\documentclass[twocolumn,aps,prl,reprint,groupedaddress]{revtex4}
\usepackage{graphicx}
\usepackage{color}
\usepackage{amsmath}

\newcommand{\be}{\begin{equation}}
\newcommand{\ee}{\end{equation}}

\newcommand{\bea}{\begin{eqnarray}}
\newcommand{\eea}{\end{eqnarray}}

\newcommand{\Li}{{\rm Li}}
\newcommand{\p}{\partial}

\newcommand{\la}{\left\langle}
\newcommand{\ra}{\right\rangle}
\newcommand{\lb}{\left[}
\newcommand{\rb}{\right]}
\newcommand{\lp}{\left(}
\newcommand{\rp}{\right)}

\renewcommand{\vec}[1]{{\bf #1}}

\newcommand{\addNB}[1]{\textcolor{magenta}{#1}}

\begin{document}
\title{Ballistic Heat Transfer and Energy Waves in an Electron System}

\author{Trung V. Phan$^1$}
\author{Justin C. W. Song$^{1,2}$}
\author{Leonid S. Levitov$^1$} 

\affiliation{$^1$ Department of Physics, Massachusetts Institute of Technology, Cambridge, Massachusetts 02139, USA}
\affiliation{$^2$ School of Engineering and Applied Sciences, Harvard University, Cambridge, Massachusetts 02138, USA}





\begin{abstract}
Materials in which heat and entropy can be transmitted by directed ballistic pulses can trigger new approaches to energy transduction in solids. We predict that a ballistic energy transfer mode, with heat propagation governed by a wave equation rather than a diffusion equation, can be realized for a thermal electron-hole plasma in graphene. 
The new behavior originates from rapid exchange of energy and momentum in particle collisions leading to energy propagation as a collective weakly-damped oscillation. Due to the electronic nature of this mode, the estimated propagation velocity can be $10^3$ times larger than that for previously studied phonon mechanisms. The energy mode is uncharged at charge neutrality, but becomes coupled to charge dynamics upon doping. This coupling can be used for all-electric excitation and detection of energy transport.
\end{abstract} 

\pacs{}
\maketitle
Can heat in a solid propagate by directed ballistic pulses?
While energy transport is typically of a diffusive character, some materials can display a very different heat transfer mode 
--- energy pulses transmitted in a collective wave-like fashion.
For such pulses the distance travelled scales linearly with the travel time, $\Delta x=v\Delta t$.
Thermal waves, called second sound, can occur in solids that
host a ``thermal liquid" of phonons.\cite{smith_jensen}
The existence of second sound requires that its frequency satisfies $\gamma_p\ll\omega\ll \gamma_N$, where  $\gamma_N$  and $\gamma_p$ are the phonon momentum-conserving (normal) and momentum-nonconserving (Umklapp) scattering rates. The rate $\gamma_N$ grows with temperature, however the rate $\gamma_p$, grows even faster. For this reason second sound, originally discovered in superfluid He,\cite{peshkov44} was observed only in a handful of solids, namely solid He, NaF and Bi.\cite{Dynes}
Second sound speed is close to $s'=s/\sqrt{3}$, where $s$ is the sound velocity,
giving values such as $20\,{\rm m/s}$ for ${}^4$He and $780\,{\rm m/s}$ for Bi. The relatively low velocity values facilitate experimental detection of second sound, yet they also limit its utility in energy transduction applications.

Can heat transfer occur at supersonic speeds? So far supersonic heat transfer has not been known in an earthly setting. 
However, theory predicts  energy and entropy waves 
in interacting systems of relativistic particles.\cite{LandauLifshitz_v6}
These  long-wavelength oscillations, sometimes called {\it cosmic sound}, propagate with very high  velocity  $c'=c/\sqrt{3}$, where $c$ is the speed of light (the quantity under square root is space dimensionality). 
Acoustic oscillations  obeying this relation underpin the modern interpretation of Cosmic Microwave Background Radiation, a relict of the ``big bang'' creation of the universe\cite{Sunyaev,Peebles}. The relation $\Delta x=c'\Delta t$ serves as a ``standard ruler" for relating cosmic length scales and distant times.

Here we predict waves analogous to cosmic sound in graphene.
Electrons in graphene behave as relativistic particles moving with velocity $v\approx 10^6\,{\rm m/s}$. 
At charge neutrality graphene hosts charge-compensated plasma with strong interactions. 
The carrier-carrier scattering in this system, under typical conditions,  is much faster than electron-lattice cooling rate and momentum relaxation due to disorder scattering, which makes graphene ideal for realizing an electronic analog of second sound. Rapid exchange of energy and momentum among colliding particles results in energy propagating as a collective excitation shared by many particles. Furthermore, since momentum is conserved alongside with energy, energy transport is characterized by inertia (finite momentum associated with energy flux). As we will see, coupled energy and momentum transport gives rise to wave-like propagation of energy with a characteristic velocity
\be\label{eq:velocity}
v'=\frac{v}{\sqrt{2}}\approx 0.71\cdot 10^6\,{\rm m/s}
,
\ee
where $v$ is the Fermi velocity. The value $v'$  is about $10^3$ times larger than the velocity for the phonon mechanism.  

\begin{figure}
\includegraphics[scale=0.33]{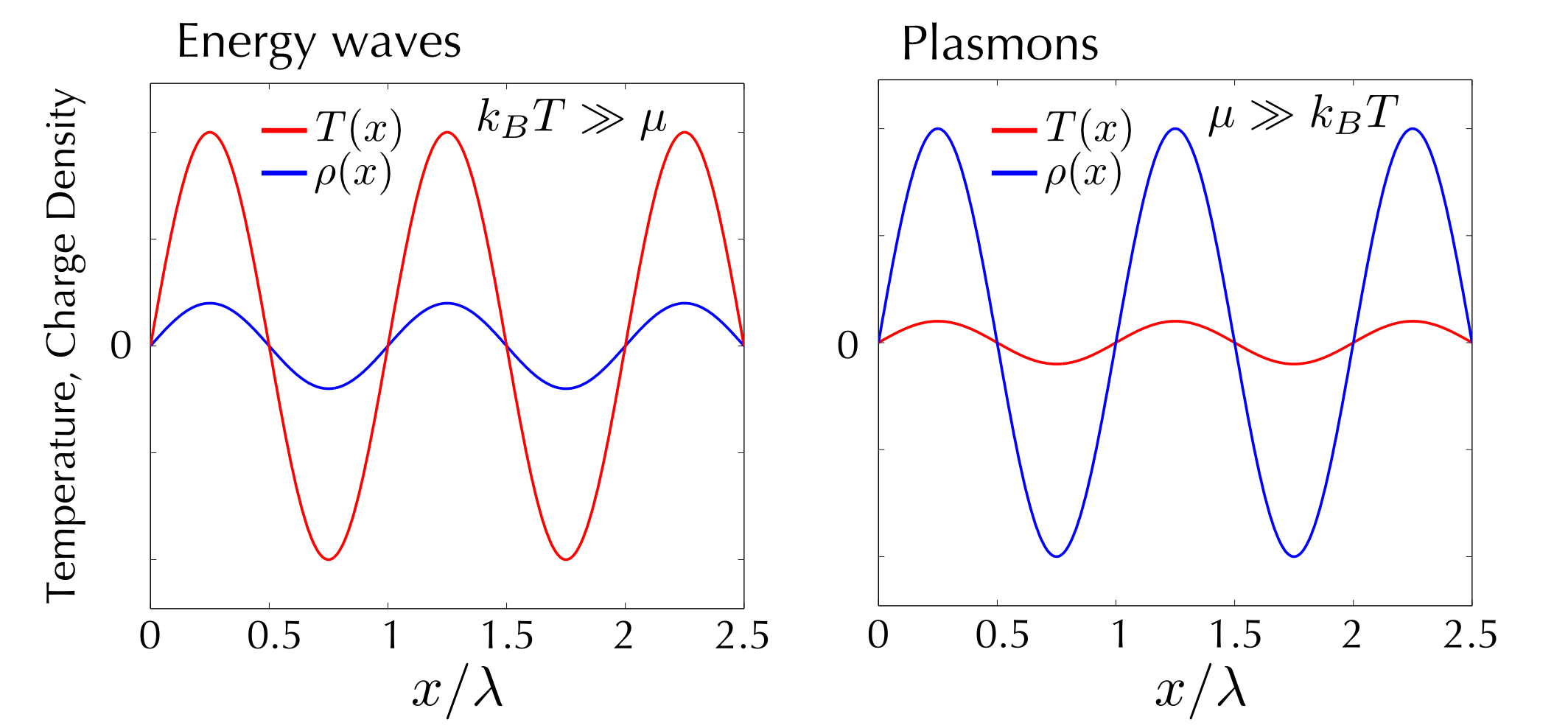}
\caption{Oscillating temperature profile that results from a collective energy
wave. Energy waves are decoupled from charge dynamics at charge neutrality, $\mu=0$,  become  charge-coupled upon doping away from neutrality, $\mu\ne0$, and eventually morph into plasmons at large $\mu$. Coupling to plasmons, Eq.(\ref{eq:dispersion_thermoplasmons}), can enable all-electric detection of energy waves.}
\label{figintro}
\vspace{-5mm}
\end{figure}

Pulse-echo measurements and standing wave resonances were used to probe phonon second sound.\cite{peshkov44,Dynes} 
For electronic second sound in graphene, gate-tunability of carrier density 
provides a new knob to probe ballistic energy transfer, having no analog 
in phonon second sound or cosmic sound. We will see that energy waves, which are fully uncharged at charge neutrality, become coupled to charge dynamics upon doping (see Fig. \ref{figintro}). Coupling to plasmons provides an all-electric way to excite and detect energy waves.
One promising approach is the recently developed spatially resolved  nanoscale probe of plasmonic standing waves.\cite{koppens,basov} Local probes can be used to excite and detect energy waves using their coupling to plasmons near charge neutrality, $k_{\rm B}T\gg \mu$. 
Alternatively, a pulse echo technique similar to that used for detecting ballistic electron resonances in carbon nanotubes\cite{mceuen} can be used.

The origin of collective thermal waves can be understood on very general grounds. 
We will first discuss cosmic sound, and then generalize to electronic waves in graphene.
The universal behavior of relativistic thermal waves follows from  the proportionality between the energy flux and momentum density of a relativistic gas, $\vec j_\epsilon=c^2\vec p$, and the relativistic pressure-energy relation, $P=\frac13W$. 
The wave equation follows under very general assumptions from energy and momentum conservation written in terms of the $4\times 4$ stress tensor:
\be\label{eq:dT=0}
\p_i T_{ij}=0
,\quad
T=  \lp \begin{array}{cc} 
W & \vec j_\epsilon \\ 
\vec j_\epsilon & \vec P \end{array}\rp
,
\ee
where $\vec P$ is a shorthand for a diagonal $3\times3$ matrix $P\delta_{\alpha\beta}$. 
Acoustic oscillations  obeying the law $c'=c/\sqrt{3}$ arise directly from the coupled dynamics of energy and momentum governed by Eq.(\ref{eq:dT=0}).

The origin of energy waves in graphene can be understood in direct analogy with cosmic sound. This is done most easily starting from conservation laws for energy and momentum of the electron system, described by the continuity equations 
\be\label{eq:continuity}
\p_tW+\nabla\vec j_{\rm q}=-\gamma_w(W-W_0)
,\quad
\p_tp_i+\nabla_j\sigma_{ij}=-\gamma_p p_i
.
\ee
Here $\vec j_{\rm q}$ is the heat current, $\sigma_{ij}$ is the stress tensor. The rates of momentum relaxation and electron-lattice cooling, $\gamma_p$ and $\gamma_w$, are introduced to account for disorder scattering and energy loss to the lattice. 
For simplicity, 
we suppressed coupling to electric field, which limits the present discussion to charge neutrality where energy and momentum oscillations are decoupled from charge oscillations. 
This coupling will be reinstated below when we discuss the effect of doping, and the relation between thermal waves and plasmons.

As we will see, the  quantities $\vec j_{\rm q}$ and  $\sigma_{ij}$
depend on momentum density and temperature in a way mimicing that in relativistic gas.
This dependence generates a cross-coupling between  the $T$ and $\vec p$ dynamics, leading to oscillations and wave propagation. Specifically, 
\be\label{eq:j,sigma_1}
\vec j_{\rm q}=v^2\vec p -\kappa\nabla T
,\quad
\sigma_{ij}=\frac12 W\delta_{ij}-\nu_1\nabla_j p_i-\nu_2\delta_{ij}\nabla_k p_k
,
\ee
were $v$ is the velocity, $W(T)\sim T^3$ is the energy density, $\kappa$ is thermal conductivity, and $\nu_{1,2}$ are viscosity parameters. The first terms in $\vec j_{\rm q}$ and  $\sigma_{ij}$ generate a nontrivial coupling of the $T$ and $\vec p$ dynamics. Since these terms do not depend on gradients, they provide a dominant contribution in the long-wavelength limit:  suppressing the terms in Eq.(\ref{eq:j,sigma_1}) first-order in gradients, we find coupled equations $\p_t W+v^2\p_j p_j=0$, $\p_t p_i+\frac12\p_i W=0$. This yields a wave equation $(\p_t^2-\frac12v^2\nabla^2)W=0$ describing energy propagation with the velocity given in Eq.(\ref{eq:velocity}). In the absence of damping, wave-like energy transport results in {\it ballistic} energy propagation, whereby distance travelled scales linearly with time, $\ell\sim v't$.

The relation between momentum density and energy flux which is key in the above analysis  can be obtained by analyzing a homogeneous macroscopic flow of Dirac particles. This is described by  a distribution
\be\label{eq_E+/-}
n(\vec p)=\frac1{e^{\beta(\epsilon_{\vec p}^{(\pm)}-\vec u\vec p-\mu)}+1}
,\quad \epsilon_{\vec p}^{(\pm)}=\pm v|\vec p|
,
\ee 
with $\vec u$ the velocity. At charge neutrality, $\mu=0$, we evaluate momentum density at first order in $\vec u$. After some algebra we find
\be\label{eq:<p>}
\la \vec p\ra=
\sum_{\vec p} \vec p\delta n(\vec p)=N\frac{9\zeta(3)p_T^3}{4\pi \hbar^2 v}\vec u
.
\ee
Here $\sum_{\vec p}... =N\sum_{\pm}\int ...\frac{d^2p}{(2\pi\hbar)^2}$, 
$N=4$ is the number of spin/valley species, and $p_T=k_{\rm B}T/v$. 
The sum is taken over both Dirac bands, $\epsilon_{\vec p}^{(\pm)}=\pm v|\vec p|$, and the quantity $\delta n(\vec p)=-(\vec u\vec p)\p f/\p \epsilon$ is a first-order variation in $\vec u$.  Next, evaluating the energy flux  at first order in $\vec u$ and comparing to Eq.(\ref{eq:<p>}), we find 
\be
\vec j_{\rm q}=\sum_{\vec p} 
\epsilon_{\vec p}^{(\pm)} \vec v\delta n(\vec p)
=N\frac{9\zeta(3)p_T^3v}{4\pi \hbar^2 }\vec u =v^2\la \vec p\ra
\ee
which determines the coupling of momentum to the heat flux, Eq.(\ref{eq:j,sigma_1}).
This relation captures the essence of the coupling between collective particle motion and energy flow in the system. 

Various damping mechanisms can be most easily assessed by analyzing solutions for Eqs.(\ref{eq:continuity}).
Taking the quantities $\vec j_{\rm q}$ and  $\sigma_{ij}$ in the form given in Eq.(\ref{eq:j,sigma_1}), and constructing plane-wave solutions,  $W,\vec p\sim e^{-i\omega t+i\vec k\vec x}$, we find the dispersion relation
\be\label{eq:dispersion}
(\omega + iD\vec k^2+i\gamma_p)(\omega + i\nu \vec k^2+i\gamma_w)=\frac{v^2}2 \vec k^2
\ee
where we defined diffusivity  $D=\kappa/C$, where $C$ is specific heat, and $\nu=\nu_1+\nu_2$. 
Eq.(\ref{eq:dispersion}) indicates that thermal waves are heavily damped at very high and very low frequencies. Damping at high frequencies is dominated by the diffusive and viscous terms, whereas damping at low frequencies is dominated by $\gamma_p$ and $\gamma_w$.

To understand the relation between various timescales, we will use parameter values estimated for pristine graphene samples which are almost defect free, such free-standing graphene.\cite{bolotin2008} 
Disorder scattering can be estimated from the measured mean free path values which reach a few microns at large doping \cite{thiti13}. Using the momentum relaxation rate square-root dependence on doping, $\gamma_p
\propto n^{-1/2}$, and extrapolating to charge neutrality, $n\sim 10^{10}\,{\rm cm^{-2}}$, gives values $\gamma_p^{-1}\sim 0.5\,{\rm ps}$. Theory predicts slow electron cooling in pristine graphene at neutrality, giving $\gamma_w^{-1}\sim T^{-2}$ with predicted values as large as $10\,{\rm ns}$ for  $T=100\, {\rm K}$\cite{bistritzer09}. Cooling times can be shortened in the presence of disorder scattering \cite{song12}, however, under any circumstances, cooling is expected to be much slower than momentum relaxation, $\gamma_w\ll \gamma_p$.  The carrier-carrier scattering which dominates the hydrodynamical regime has little effect on cooling. 

Refs.\cite{kashuba08,fritz08} estimate 
the  carrier-carrier scattering rate as $\gamma_N\approx A \alpha^2 k_{\rm B}T/\hbar$, where  $\alpha$ is the interaction strength. For $T=100\,{\rm K}$, approximating the prefactor as $A\approx 1$\cite{kashuba08,fritz08}, we obtain characteristic times, $\gamma_N^{-1} \approx 70 \, {\rm fs}$, which are much shorter than the values $\gamma_p^{-1}$ and $\gamma_w^{-1}$ estimated above. 
The inequalities $\gamma_w\ll \gamma_N$, $\gamma_p\ll \gamma_N$ justify our hydrodynamical description of transport. 
Estimating the diffusion constant and viscosity, $D,\,\nu\approx\frac12 v^2/\gamma_N$, and comparing to the dispersion relation, Eq.(\ref{eq:dispersion}), we see that damping is weak in the range of frequencies defined as
\[
\gamma_p\lesssim \omega\lesssim \gamma_N
\]
Using the values estimated above, gives frequencies $f=\omega/2\pi$ in the low THz range, $0.1\,{\rm THz}\lesssim f \lesssim 
3\,{\rm THz}$. 

Next we will analyze the effect of finite doping. As we will see, the thermal waves morph  into plasma waves 
upon doping away from neutrality. We will employ a microscopic approach based on the quantum kinetic equation, which will allow us to justify our hydrodynamical treatment
by a microscopic analysis. The kinetic equation for electron distribution reads
\be\label{eq:kin_eq}
\lp \p_t+\vec v\nabla_{\vec x}+e\vec E\nabla_{\vec p}\rp n(\vec x,\vec p,t)=I^N    +I^{\rm el-ph}+I^{\rm dis} 
,
\ee
where $\vec E$ is the electric field, and $I^N$,  $I^{\rm el-ph}$,  $I^{\rm dis}$ describe (normal) two-particle collisions, electron-phonon collisions and scattering by disorder. The electric field can be either extrinsic (imposed externally) or intrinsic, arising due to long-wavelength charge fluctuations (as will be the case in our analysis of plasmons).

Collective transport arises in the regime dominated by normal collisions, when the processes $I^{\rm el-ph}$ and  $I^{\rm dis}$ are negligible. 
The quantity $I^N$ is given by the standard expression which accounts for energy and momentum conservation in two-particle collisions,\cite{Lifshitz_Pitaevskii}
\bea \nonumber
 && I^N[n]=\!\!\!\sum_{\vec p_1,\vec p',\vec p_1'}w_{\vec p',\vec p_1'\to \vec p,\vec p_1}\lb
(1-n(\vec p))(1-n(\vec p_1))\right.
\\ \nonumber
&& \left. \times n(\vec p')n(\vec p_1') -(1-n(\vec p'))(1-n(\vec p_1'))n(\vec p)n(\vec p_1)
\rb
\\ \label{eq:W123}
&& \times \delta(\vec p+\vec p_1-\vec p'-\vec p'_1)\delta(\epsilon_{\vec p}+\epsilon_{\vec p_1}-\epsilon_{\vec p'}-\epsilon_{\vec p'_1})
.
\eea
The transition rate $w$ is given by the electron scattering vertex which includes coherence factors, e.g. Born approximation yields $w_{\vec p',\vec p_1'\to \vec p,\vec p_1}=\frac{2\pi}{\hbar} |V_{\vec p,\vec p_1,\vec p'\vec p'_1}|^2$. 

In the long-wavelength limit we can analyze solutions of the kinetic equation, Eq.(\ref{eq:kin_eq}), perturbatively for 
a weak inhomogeneity. Setting the left-hand side of Eq.(\ref{eq:kin_eq}) to zero (electric field becomes small in the long-wavelength limit), we find that the kinetic equation is approximately solved by zero modes of the collision integral. The zero-mode equation $I^N [n]=0$ can be solved by taking into account the energy-conserving and momentum-conserving character of the collision operator. A standard Boltzmannesque reasoning gives
a general solution of the form\cite{Lifshitz_Pitaevskii}
\be\label{eq:n0_general}
n(\vec x,\vec p,t)=\frac1{e^{\beta(\vec x,t) (\epsilon_{\vec p}^{(\pm)}-\vec p \vec u(\vec x,t)-\mu(\vec x,t)}+1}
.
\ee
This equation determines the $\vec p$ dependence of particle distribution, leaving the dependence of the quantities $\vec u$, $\beta$ and $\mu$  on position and time unspecified. 
Here $\vec u(\vec x,t)$ is the hydrodynamical velocity describing collective motion of the e-h plasma, $\beta(\vec x,t)$ describes temperature variation, and $\mu(\vec x,t)$ describes the local chemical potential deviation from equilibrium. Accounting for variations in $\mu$ is inessential at charge neutrality, where energy and momentum oscillations occur at $\mu(\vec x,t)=0$. However, as we will see, $\mu$ must be included in the full analysis 
to achieve a unified description of both the undoped and doped regime. In particular, plasma oscillations in the doped regime are associated with a time-varying $\mu$.

The hydrodynamic description of energy and momentum transport can be obtained 
directly from conservation laws for particle density, energy and momentum. Specifically, we consider deviations in these quantities from equilibrium, $\delta N=\sum_{\vec p} \delta n(\vec p)$, $\delta W=\sum_{\vec p}\epsilon_{\vec p}^{(\pm)}\delta n(\vec p)$, $\la \delta \vec p\ra =\sum_{\vec p}\vec p \delta n(\vec p)$. 
Integrating Eq.(\ref{eq:kin_eq}) over $\vec p$ yields the continuity equation for particle number. 
Multiplying Eq.(\ref{eq:kin_eq}) by $\epsilon_{\vec p}^{(\pm)}$ (or, by $\vec p$), integrating, and accounting for energy (momentum) conservation by normal collisions, yields equations for energy and momentum transport, 
\bea\label{eq:conservation_laws_NE}
&&\p_t \delta N +\nabla \vec j_N =0,\quad \p_t \delta W +\nabla \vec j_{\rm q}=0,
\\\label{eq:conservation_laws_P}
&&\p_t \la \delta p_i\ra +\nabla_j \sigma_{ij}-n_0 eE_i=0
.
\eea
Here $\vec j_N =\la \vec v \ra$ and $\vec j_{\rm q}=\la \vec v \epsilon_{\vec p}\ra$ are particle and energy currents, $\sigma_{ij}=\la v_i p_j\ra$ is the stress tensor, and $n_0$ is carrier density. In the interest of brevity we will denote averaging over the distribution variation $\delta n$ by angular brackets, $\la{\cal A}\ra=\sum_{\vec p}{\cal A} \delta n(\vec p)$, and suppress the $\pm$ superscript of $\epsilon_{\vec p}$. Here we used the continuity equation for $\delta N$ to simplify the equation for $\delta W$ by dropping $\mu$:
$\la  \epsilon_{\vec p}-\mu\ra\to \la \epsilon_{\vec p}\ra$, $\la  \vec v (\epsilon_{\vec p}-\mu)\ra\to \la  \vec v \epsilon_{\vec p}\ra$.  

The hydrodynamic equations at first order in deviations from equlibrium can be obtained by expanding the expression in Eq.(\ref{eq:n0_general}) as
\be
\delta n=\frac{\p n}{\p \beta}\delta\beta(\vec x,t)+\frac{\p n}{\p \mu}\delta\mu(\vec x,t)+\frac{\p n}{\p u_i}\delta u_i(\vec x,t)
,
\ee
plugging in the conservation laws and integrating over $\vec p$. In doing so it will be convenient to combine the first two terms, which are isotropic in $\vec p$, and denote them as $\delta_1 n$. The last term, which has angular dependence of the form $\vec p\vec u$, will be denoted as $\delta_u n$. Eq.(\ref{eq:n0_general}) gives
\be\label{eq_delta_n}
\delta_1 n=\lp \frac{\epsilon-\mu}{\beta}\delta\beta-\delta\mu\rp\frac{\p n}{\p \epsilon}
,\quad
\delta_u n=-\vec p\delta\vec u \frac{\p n}{\p \epsilon}
\ee
Combining these expressions with the conservation laws, and taking into account the angular dependence in Eq.(\ref{eq_delta_n}), we see that the quantities $\delta N$, $\delta W$ and $\sigma_{ij}$ are expressed through $\delta_1 n$, whereas $\vec j_N$, $\vec j_{\rm q}$ and $\la \delta p_i\ra$ are expressed through $\delta_u n$. Then, with angular averaging performed via $\la \vec v\delta_u n\ra=-\frac12\vec u\la\epsilon_{\vec p}\ra$, the conservation laws for particle number and energy, Eq.(\ref{eq:conservation_laws_NE}), become
\be\label{eq:_AB}
\p_t\la \delta_1 n\ra=A \nabla\vec u
,\quad
\p_t\la \epsilon_{\vec p}\delta_1 n\ra=B \nabla\vec u
,
\ee
where $A=\la\frac{\epsilon_{\vec p}}2\frac{\p n}{\p \epsilon}\ra$, $B=\la\frac{\epsilon_{\vec p}^2}2\frac{\p n}{\p \epsilon}\ra$.
We note parenthetically that if $\delta \mu(\vec x,t)$ were not treated as an independent hydrodynamical variable, the equations in Eq.(\ref{eq:_AB}) would become incompatible with each other. 
Consistency is assured by $\delta \mu$ varying independently of $\delta \beta$ and $\delta\vec u$. 

Lastly, we consider the momentum transport equation, Eq.(\ref{eq:conservation_laws_P}). The stress tensor, after angular averaging performed as above, can be written as 
$\sigma_{ij}=\frac12\delta_{ij}\la\epsilon_{\vec p}\delta_1 n\ra$.
Expressing the electric field through spatial variation of particle density, we have
$\vec E=-\nabla \int \frac{e}{|\vec x-\vec x'|} \la \delta_1 n_{\vec x'}\ra d^2 x'$.
We note that $\sigma_{ij}$ and $\vec E$ are expressed through the same quantities as those appearing under time derivatives in Eq.(\ref{eq:_AB}). With the help of this observation, we can write the time derivative of Eq.(\ref{eq:conservation_laws_P}) as
\be\label{eq_partial^2}
\p_t^2 \la \delta \vec p\ra +\frac12 B\nabla^2\vec u=-n_0\nabla \int \frac{e^2}{|\vec x-\vec x'|} A\nabla\vec u(\vec x')d^2 x'
.
\ee
Finally, we express $ \la \delta \vec p\ra$ through $\vec u$ by writing 
\be
\la \delta  \vec p\ra= \la \vec p\delta_u n\ra=-\la\frac{\vec p^2}2\frac{\p n}{\p \epsilon}\ra\vec u=-\frac{B}{v^2}\vec u
.
\ee
Plugging this in 
Eq.(\ref{eq_partial^2}), and passing to Fourier harmonics, $\vec u(\vec x,t)\sim e^{i\vec k\vec x-i\omega t}$, we obtain the dispersion relation describing charge-coupled thermal waves:
\be\label{eq:dispersion_thermoplasmons}
\omega^2=\frac{v^2}2\vec k^2+ 2\pi e^2 \lambda n_0 v^2|\vec k|
,\quad
\lambda=\frac{A}{B}
.
\ee
The plasmonic correction (second term) vanishes at charge neutrality, $\mu=0$, and is small near it. 
Numerical analysis of Eq.(\ref{eq:dispersion_thermoplasmons}) 
can be done using the integral $\int_0^\infty \frac{t^{s-1}}{e^t/z+1}dt=-\Gamma(s)\Li_s(-z)$, where $\Li_s$ is the polylogarithm function and $z=e^{\beta\mu}$. We find
\be
\lambda n_0=-\alpha\frac{\lp \Li_2(-z)-\Li_2(-1/z)\rp^2}{ \Li_3(-z)+\Li_3(-1/z)}\approx \frac{16\ln^2 2\beta\mu^2}{9\pi\zeta(3)\hbar^2v^2}
,
\ee
with the last equation valid near charge neutrality $|\mu|\ll k_{\rm B}T$ [here $\alpha=\frac{2}{3\pi\hbar^2v^2\beta}$]. 
The last term in Eq.(\ref{eq:dispersion_thermoplasmons}) becomes dominant far from neutrality, $|\mu|\gg k_{\rm B}T$. In this limit, we reproduce the standard plasmon dispersion.\cite{hwang2007} Charge coupling that is nonzero but small near charge neutrality, Eq.(\ref{eq:dispersion_thermoplasmons}), provides a convenient tool for an all-electric excitation and detection of energy waves. 

In summary, graphene can host new collective modes, namely long-wavelength energy oscillations that propagate as weakly  damped waves. Energy waves exist in the hydrodynamical frequency range, $\omega<\gamma_N$. This condition sets such waves apart from 
various types of collisionless collective modes proposed at charge neutrality, which occur at $\omega\ge\gamma_N$.\cite{vafek2006,mishchenko2008,dassarma2013} The electronic nature of energy waves ensures their high propagation velocity, which can be $10^3$ times larger than the highest values known for the phonon mechanism. 
Directed ballistic energy  pulses enabled by electronic second sound open the door 
to achieving high-speed energy transduction in solids.

We thank
A. L. Efros, V. F. Gantmakher, P. L. McEuen, M. Yu. Reizer and D. Son for useful discussions.


\vspace{-4mm}
\begin{thebibliography}{99}
\bibitem{smith_jensen}
H. Smith, H. H. Jensen,  (1989). Transport Phenomena. Oxford University Press. 
Sec. 4.3: Second Sound. 
\bibitem{peshkov44} V. Peshkov,  J. Phys. (Moscow) 8, 381 (1944). 
\bibitem{Dynes} V. Narayanamurti, R. Dynes, 
Phys. Rev. Lett. 28, 1461 (1972). 
\bibitem{LandauLifshitz_v6} L. D. Landau and E. M. Lifshitz, Fluid Mechanics. Vol. 6 (Butterworth-Heinemann, 1987). 
\bibitem{Sunyaev} R. Sunyaev and Ya.B. Zeldovich, Astrophysics and Space Science. 7, 3 (1970). 
\bibitem{Peebles} P. J. E. Peebles and J. T. Yu, The Astrophysical Journal. 162, 815 (1970). 
\bibitem{koppens} J. Chen et al. 
Nature (2012) doi:10.1038/nature11254
\bibitem{basov}  Z. Fei et al. 
Nature (2012) doi:10.1038/nature11253
\bibitem{mceuen}
Z. Zhong, N. M. Gabor, J. E. Sharping, A. L. Gaeta, and P. L. McEuen, Nature Nano. 3, 201-205 (2008). 
\bibitem{bolotin2008} K. I. Bolotin et al. 
Solid State Comm. {\bf 146}, 351-355 (2008).
\bibitem{thiti13} T. Taychatanapat, K. Watanabe, T. Taniguchi, P. Jarillo-Herrero, Nature Physics, {\bf 9} 225 (2013).
\bibitem{bistritzer09} R. Bistritzer, A. H. Macdonald, Phys. Rev. Lett., {\bf 102} 206410 (2009).
\bibitem{song12} J. C. W. Song, M. Y. Reizer, L. S. Levitov, Phys. Rev. Lett., {\bf 109} 106602 (2012).
\bibitem{kashuba08} A. B. Kashuba, Phys. Rev. B, {\bf 78} 085415 (2008).
\bibitem{fritz08} L. Fritz, J. Schmalian, M. M\"uller, and S. Sachdev, Phys. Rev. B, {\bf 78} 085416 (2008).

\bibitem{Lifshitz_Pitaevskii} E. M. Lifshitz, L. P. Pitaevskii, Physical Kinetics, Vol. 10 (Butterworth-Heinemann, 1981).
\bibitem{gurzhi63} R. N. Gurzhi, Pis'ma Zh. Eksp. Teor. Fiz. 44, 771 (1963) [JETP Lett. 17, 521 (1963)]; Usp. Fiz. Nauk 94, 689 (1968) [Sov. Phys. Usp. 11, 255 (1968)].

\bibitem{gurzhi95} R. N. Gurzhi, A. N. Kalinenko, and A. I. Kopeliovich, Phys. Rev. Lett. 74, 3872-3875 (1995)

\bibitem{vafek2006} O. Vafek, Phys. Rev. Lett. 97, 266406 (2006).
\bibitem{mishchenko2008} S. Gangadharaiah, A. M. Farid, and E. G. Mishchenko, Phys. Rev. Lett. 100, 166802 (2008).

\bibitem{dassarma2013} S. Das Sarma, Q. Li, arXiv:1305.3605 
\bibitem{hwang2007} E. H. Hwang and S. Das Sarma, Phys. Rev. B 75, 205418 (2007). 

\end{thebibliography}
\end{document}